\documentclass[pra,aps,amsmath,amssymb,amsfonts,nofootinbib,superscriptaddress,10pt,onecolumn]{revtex4-2}
\usepackage{bm,mathrsfs}
\usepackage{graphicx}
\usepackage{epsfig}
\usepackage{verbatim}
\usepackage[percent]{overpic}
\usepackage{subcaption}

\usepackage[margin=2cm]{geometry}
\usepackage[colorlinks,breaklinks,linkcolor=blue,anchorcolor=blue,citecolor=blue,urlcolor=blue,bookmarks=false]{hyperref}
\newcommand{\ket}[1]{|#1\rangle}
\newcommand{\bra}[1]{\langle#1|}

\begin{document}
	
	\title[Short Title]{Deterministic interconversion of the Greenberger–Horne–Zeilinger state and Knill–Laflamme–Milburn state via Lie-transform-based pulse design in Rydberg atoms}
	
	\author{J. P. Wang}
	\affiliation{College of Physical Science and Technology, Bohai University, Jinzhou 121013, China}
	\author{Y. Q. Ji}
	\email[E-mail:]{jiyanqiang@qymail.bhu.edu.cn}
	\affiliation{College of Physical Science and Technology, Bohai University, Jinzhou 121013, China}
	\author{L. P. Yang}
	\affiliation{College of Physical Science and Technology, Bohai University, Jinzhou 121013, China}
	\author{C. Q. Wang}
	\affiliation{College of Physical Science and Technology, Bohai University, Jinzhou 121013, China}
	\author{L. Dong}
	\affiliation{College of Physical Science and Technology, Bohai University, Jinzhou 121013, China}
	\author{X. M. Xiu}
	\email[E-mail:]{xiuxiaomingdl@126.com}
	\affiliation{College of Physical Science and Technology, Bohai University, Jinzhou 121013, China}

	\begin{abstract}
		Conversion between different types of entangled states is an interesting problem in quantum mechanics. But research on the conversion between Greenberger-Horne-Zeilinger (GHZ) state and Knill-Laflamme-Milburn (KLM) state in atomic system has not been reported. In this paper, we propose a scheme to realize the interconversion (one-step) between GHZ state and KLM state with Rydberg atoms. By utilizing Rydberg-mediated interactions, we simplify the system. By combining Lie-transform-based pulse design, the evolution path is built up to realize interconversion of GHZ state and KLM state. The numerical simulation result shows that the present scheme is robust against decoherence and operational imperfection. 
	\end{abstract}
	
	\maketitle

\section{Introduction}\label{sec1}
Quantum entanglement~\cite{Horodecki2009,vedral2014quantum,mintert2005measures} is one of the most fascinating phenomena in quantum mechanics. Entangled states are useful for exploring non-locality, bolstering quantum mechanics against theories of local hidden variables. Moreover, entangled states also are key resource for quantum information processing tasks~\cite{monroe2002quantum,blais2020quantum,xavier2020quantum}. It is widely used in quantum teleportation~\cite{qt1,qt2,santos2016shortcut}, quantum key distribution~\cite{qd1,qd2,shor2000simple}, quantum cryptography~\cite{vedral2014quantum,bernstein2017post,pirandola2020advances,RevModPhys.94.025008},   quantum computing~\cite{qc1,qc3} and so on.  So far, many theoretical and experimental schemes have been proposed for preparing and converting entangled states~\cite{c1,Song2012,shao2023,liu2022,zhang2024fast,ji2017conversion,wu2017superadiabatic,Gujarati2018,Ji2019,Kang2019,shen2019conversion}. As we know, each type of entangled state has its own characteristics, therefore, exploring the conversion between different types of entangled states has become a fascinating research direction in quantum mechanics.

On the other hand, neutral atoms are regarded as ideal candidates for quantum information processing due to stabilized atomic hyperfine energy states and extremely large dipole moments, which are particularly suitable for encoding logic quantum bits~\cite{Saffman2009,adams2019rydberg,shiQuantumLogicEntanglement2022,yang2024fast,yu2022multiqubit}. When an atom is excited to the high Rydberg state, the strong dipole-dipole interaction or van der Waals interaction will significantly shift its surrounding atomic energy levels of Rydberg states, this leads to the emergent energy level shift term in the Hamiltonian. Mathematically, it is because of this energy level shift term that a series of interesting physical phenomena arise, such as the Rydberg blocking effect~\cite{gaetan2009observation,urban2009observation,Shao2017,chen2018fast} and the Rydberg anti-blocking effect~\cite{2017singlet,PhysRevLett.104.013001,su2020rydberg,Su2021Dipole,shao2020s}. In addition, it can be a very interesting tool to simplify the dynamics of systems, as well as, provides researchers with brand new idea. There are a lot of excellent studies to realize  conversions between  difference entangled state  using neutral atoms~\cite{Song2012,c1,shao2023,liu2023fast,PhysRevA.103.032427,nauth2022quantum}. It is worth noting that Zheng $et~al.$ \cite{c1} proposed a protocol for the one-step interconversion between GHZ state and the W state with deterministic success probabilities by utilizing Rydberg-mechanism to structure nonlocal operation. It has been demonstrated that the two types of entangled states cannot be interconverted only by local operations and classical communication. We consider that since a deterministic interconversion between GHZ and W states can be realized in one step by utilizing Rydberg-mechanism to structure nonlocal operation~\cite{c1}, it might also be possible to realize the deterministic conversion between other entangled states in one step.

We focus on another type of  Knill-Laflamme-Milburn (KLM) entangled state ~ \cite{KLM}. The KLM state was introduced mainly in the linear optical quantum computing scheme proposed by Knill, Laflamme and Milburn. It can effectively reduce the operation error rate and improve the implementation efficiency of quantum algorithms. Subsequently, not limited to optical systems, Popescu  \cite{popescu2007knill} demonstrated that the atomic KLM can also be used for computation.
Since then, considerable efforts have been invested in preparing KLM-type quantum entanglement using various physical platforms~\cite{Lemr2010Experimental,okamoto2011realization,cheng2012generation,ji2020preparation,shen2018multiphoton,li2018engineering,zheng2021Fast}, such as linear optics, atom-cavity quantum electrodynamics, nonlinear cross-Kerr mediums, and artificial atoms. The equal-weight $n$-qubit KLM state is given by
\begin{equation}
	\ket{\mathrm{KLM}}
	= \alpha_n \sum_{k=0}^{n}
	\ket{1}^{\otimes k} \otimes \ket{0}^{\otimes(n-k)},
\end{equation}
where $\alpha_n = 1/\sqrt{n+1}$ is the normalization factor.

After extensive literature review, although many schemes have been proposed for directly preparing GHZ state or KLM state in atomic system~\cite{Gujarati2018,ghz1,ghz2,klm1,li2018engineering,zheng2021Fast} and a  scheme for converting GHZ state and KLM state in optical systems~\cite{shen2019conversion}, we found that research on the conversion between GHZ state and KLM state in neutral atomic systems has not been reported. In the present work, inspired by the previous schemes, we propose a scheme to deterministic implement  interconversion between GHZ state and KLM state with one-step. The scheme's physical model features three neutral atoms with Rydberg capabilities, positioned in a triangular configuration, the original Hamiltonian is simplified to an effective four-energy system through a meticulously crafted detunings and approximation technique. Then, we combine the effective Hamiltonian and Lie-transform-based pulse design~\cite{martinez2014shortcuts,kang2019pulse} to realize the interconversion of GHZ state and KLM state.

The paper is organized as follows. In Section {.} 2, we introduce the physical model and simplify the system to an effective four energy level system. In Section {.} 3, we briefly review the methods for  Lie-transform-based pulse design, subsequently, we combine effective Hamiltonian and Lie-transform-based pulse design to structure control pulses to realize the interconversion between GHZ state and KLM state. Nextly, in Section {.} 4 , we give the numerical simulations and analyses for the scheme, and further discuss the robustness about a number of imperfections. At last, a summary is given in Section {.} 5. 

\section{Physical model}\label{sec2}
\begin{figure*}[htbp]
	\centering
        \scalebox{0.6}{\includegraphics{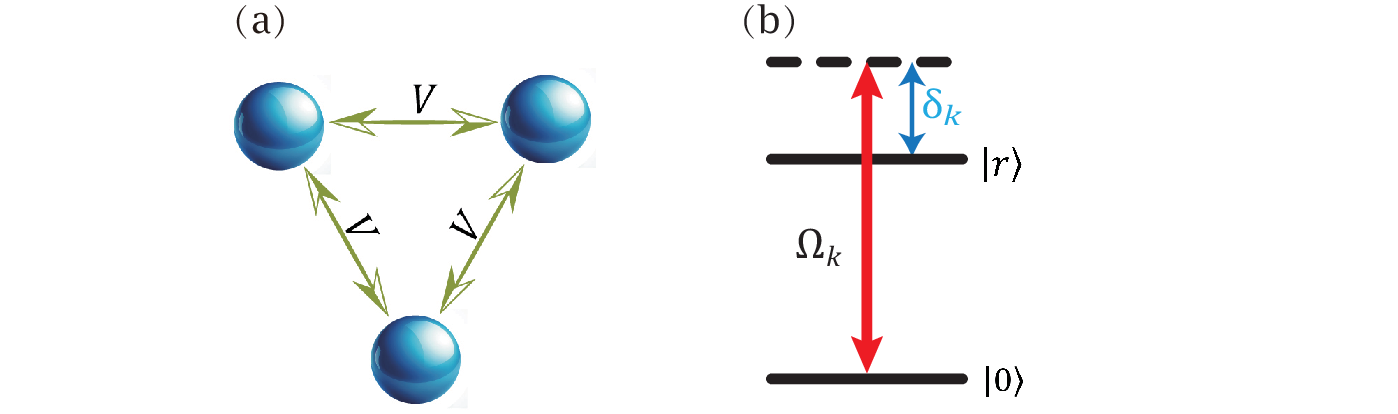}}
\caption{(a)Schematic illustration of system. Three identical neutral atoms confined within three separate microscopic dipole traps, arranged in a triangular configuration. (b)The energy levels configuration for the single atom.}    
\label{fig1}
\end{figure*}

We consider three same neutral atoms confined within three distinct microscopic dipole traps, and arranged in a triangular configuration as show in Fig.\ref{fig1}(a), $V$ represents the van der Waals or dipole–dipole interaction between different atoms that are in Rydberg states. The energy levels of the three neutral atoms with Rydberg states as shown in Fig.\ref{fig1}(b), each atomic level includes a ground states $\ket{0}$ and a Rydberg state $\ket{r}$.  The transition $\ket{0}_k\leftrightarrow\ket{r}_k$ is driven by classical  laser fields with  Rabi frequencies $\Omega_k$, whose detunings are $\delta_k$, where $k$ denotes the $k$th atom in the physical model. In the interaction picture, the  Hamiltonian of the physical model takes the form$(\hbar=1)$
\begin{align}\label{eq1}
	H(t) =& \sum_{k=1}^{3} \Omega_k  {\rm e}^{- {\rm i }\delta_k t} \ket{r}_k\bra{0} +\mathrm{H.c.}+V\sum_{k=2}^{3}\sum_{k'=1}^{k-1} \ket{rr} _{k' k}\bra{rr}.
\end{align}

To improve understanding and facilitate discussion of the physical model, we rewrite the Hamiltonian in triatomic basis vectors
{
\begin{align}\label{eq2}
	H =&\Omega_1  {\rm e}^{- {\rm i }\delta_1 t} \ket{r00}\bra{000}+\Omega_2  {\rm e}^{- {\rm i }\delta_2 t}\ket{0r0}\bra{000}+\Omega_3  {\rm e}^{- {\rm i}\delta_3 t}\ket{00r}\bra{000}  +\Omega_2  {\rm e}^{- {\rm i}\delta_2 t}\ket{rr0}\bra{r00}\cr
	&+\Omega_3  {\rm e}^{- {\rm i}\delta_3 t}\ket{r0r}\bra{r00}+\Omega_1  {\rm e}^{- {\rm i}\delta_1 t} \ket{rr0}\bra{0r0}+\Omega_3  {\rm e}^{- {\rm i}\delta_3 t} \ket{0rr}\bra{0r0}+\Omega_1  {\rm e}^{- {\rm i}\delta_1 t} \ket{r0r}\bra{00r}\cr
	&+\Omega_2  {\rm e}^{- {\rm i}\delta_2 t}\ket{0rr}\bra{00r}+\Omega_3  {\rm e}^{- {\rm i}\delta_3 t}\ket{rrr}\bra{rr0}+\Omega_2  {\rm e}^{- {\rm i}\delta_2 t}\ket{rrr}\bra{r0r}+\Omega_1  {\rm e}^{- {\rm i}\delta_1 t}\ket{rrr}\bra{0rr}+\rm{H.c.}\cr
	&+V\ket{rr0}\bra{rr0}+V\ket{r0r}\bra{r0r}+V\ket{0rr}\bra{0rr}+3V\ket{rrr}\bra{rrr}.
\end{align}
}
From Eq.(\ref{eq2}), we can easily observe the energy level shifts caused by Rydberg states, this energy level shift is 
\begin{align}  
	H_V=&V\ket{rr0}\bra{rr0}+V\ket{r0r}\bra{r0r}+V\ket{0rr}\bra{0rr}+3V\ket{rrr}\bra{rrr}.
\end{align}
Next, we define a rotating frame with a unitary operator $\mathcal{R}=\exp{( {-{\rm i}}H_Vt)}$, and move into the new picture through
\begin{align}\label{eq3}
	H'=\mathcal{R}^\dag H \mathcal{R}  {-{\rm i}}{\mathcal{R}^\dag} \Dot{\mathcal{R}}.
\end{align}
So, the Hamiltonian (\ref{eq2}) in new picture becomes
{
\begin{align}\label{eq4}
	H' =&\Omega_1  {\rm e}^{- {\rm i}\delta_1 t} \ket{r00}\bra{000}+\Omega_2  {\rm e}^{- {\rm i}\delta_2 t}\ket{0r0}\bra{000}+\Omega_3  {\rm e}^{- {\rm i}\delta_3 t}\ket{00r}\bra{000} +\Omega_2  {\rm e}^{ {-{\rm i}}(\delta_2-V) t}\ket{rr0}\bra{r00}\cr    
	&+\Omega_3  {\rm e}^{ {-{\rm i}}(\delta_3-V) t}\ket{r0r}\bra{r00}+\Omega_1  {\rm e}^{ {-{\rm i}}(\delta_1-V) t} \ket{rr0}\bra{0r0}+\Omega_3  {\rm e}^{ {-{\rm i}}(\delta_3-V) t} \ket{0rr}\bra{0r0}\cr    
	&+\Omega_1  {\rm e}^{ {-{\rm i}}(\delta_1-V) t} \ket{r0r}\bra{00r}+\Omega_2  {\rm e}^{ {-{\rm i}}(\delta_2-V) t}\ket{0rr}\bra{00r}+\Omega_3  {\rm e}^{ {-{\rm i}}(\delta_3-2V) t}\ket{rrr}\bra{rr0}\cr    
	&+\Omega_2  {\rm e}^{ {-{\rm i}}(\delta_2-2V) t}\ket{rrr}\bra{r0r}+\Omega_1  {\rm e}^{ {-{\rm i}}(\delta_1-2V) t}\ket{rrr}\bra{0rr} 
	+\rm{H.c.}
\end{align}
}
If we encode the logical `0' with ground state $\ket{0}$ and use the Rydberg state $\ket{r}$ to encode logical `1', the three-particle KLM state and GHZ state has the following form
\begin{align}
	\ket{ {\rm KLM}}=&\frac{1}{2}(\ket{000}+\ket{r00}+\ket{rr0}+\ket{rrr}),
\end{align}
and 
\begin{align}
	\ket{ {\rm GHZ}}=&\frac{1}{\sqrt{2}}(\ket{000}+\ket{rrr}).   
\end{align}
We considered that if constructing an evolutionary space within $ \{ \ket{000}, \ket{r00},\ket{rr0}, \ket{rrr} \}$ from Eq.(\ref{eq4}), it may accomplish the preparation of GHZ state and KLM  state, as well as enable their interconversion. Based on this consideration, we set the detuning of the Rabi frequencies are $\delta_1=0$, $\delta_2=V$, $\delta_3=2V$, respectively, then the Hamiltonian (\ref{eq4}) becomes 
{
\begin{align}\label{eq5}
	H'=&\Omega_1 \ket{r00}\bra{000}+\Omega_2  {\rm e}^{ {-{\rm i}}V t}\ket{0r0}\bra{000}+\Omega_3  {\rm e}^{ {-{\rm i}}2V t}\ket{00r}\bra{000} +\Omega_2 \ket{rr0}\bra{r00}\cr    
	&+\Omega_3  {\rm e}^{ {-{\rm i}}V t}\ket{r0r}\bra{r00}+\Omega_1  {\rm e}^{ {{\rm i}}V t} \ket{rr0}\bra{0r0}+\Omega_3  {\rm e}^{ {-{\rm i}}V t} \ket{0rr}\bra{0r0}+\Omega_1  {\rm e}^{ {{\rm i}}V t} \ket{r0r}\bra{00r}\cr    
	&+\Omega_2 \ket{0rr}\bra{00r}+\Omega_3 \ket{rrr}\bra{rr0}+\Omega_2  {\rm e}^{ {{\rm i}}V t}\ket{rrr}\bra{r0r}+\Omega_1  {\rm e}^{ {{\rm i}}2V t}\ket{rrr}\bra{0rr} +\rm{H.c.}
\end{align}
}
Through neglecting highly frequent oscillations under the condition $V\gg \rm{Max}\{\Omega_1,\Omega_2,\Omega_3\}$, the effective Hamiltonian is simplified as
\begin{align} \label{eq6}
	H'_{ {\rm eff}} =& \Omega_1  \ket{r00}\bra{000}+\Omega_2 \ket{rr0}\bra{r00}+\Omega_2 \ket{0rr}\bra{00r}+\Omega_3 \ket{rrr}\bra{rr0} +\rm{H.c.}
\end{align}
From Eq.(\ref{eq6}), we can find that if the initial state belongs to the  evolutionary subspace \{$\ket{000}$, $\ket{r00}$, $\ket{rr0}$, $\ket{rrr}$\}, the evolution of the system will remain within this evolutionary subspace. That is to say, in this case, the term $\Omega_2 \ket{0rr}\bra{00r}$ and its Hermitian conjugate are decoupled in effective Hamiltonian(\ref{eq6}), so  the effective Hamiltonian becomes
\begin{align}
	H_{ {\rm eff}}= &\Omega_1 \ket{r00}\bra{000}+\Omega_2 \ket{rr0}\bra{r00}\
	+\Omega_3 \ket{rrr}\bra{rr0}+\rm{H.c.}.\label{eq9}
\end{align}

\section{Control field design through inverse Hamiltonian engineering based on Lie transforms}\label{sec3}
\subsection{Lie-transforms-based pulse design}
First of all, we briefly describe the Lie-transforms-based pulse design~\cite{kang2019pulse}, for a quantum system with time-dependent interaction, assuming the Hamiltonian is expressed as follows
\begin{align}
	\mathcal{H}(t)=\sum_{j=1}^N\lambda_j(t)A_j,
\end{align}
where the parameters $\lambda_j(t) $ are time-dependent. Additionally, $\{A_j\}$ is a set of Hermitian generators spanning a Lie algebra and $\{A_j\}$ as the basis of an $N$-dimensional Hilbert space $\mathcal{A}$. A set of Lie transforms $\{\mathscr{L}_j\}$ is defined and each function 
$\mathscr{L}_j$ acting on any vector $A\in\mathcal{A}$ is
\begin{align}
	\mathscr{L}_j(A)&= {\rm e}^{ {\rm i}\theta_j(t)A_j}A {\rm e}^{- {\rm i}\theta_j(t)A_j},
\end{align}
where parameter $\theta_j$ is time-dependent and designed. Since the rotation $ {\rm e}^{ {\rm i}\theta_j (t )A_j}$ can transform the Hamiltonian from a picture to another picture, it define a picture transform $\mathscr{P}_j$ acting on $\mathcal{H}(t)$ as
\begin{align}
	\mathscr{P}_j(\mathcal{H})= {\rm e}^{ {\rm i}\theta_j(t)A_j}\mathcal{H}(t) {\rm e}^{- {\rm i}\theta_j(t)A_j}- {\rm i} {\rm e}^{ {\rm i}\theta_j(t)A_j}\frac{ {\rm d}}{ {\rm d}t}\left( {\rm e}^{- {\rm i}\theta_j(t)A_j}\right) 
	=\mathscr{L}_j(\mathcal{H}) - \dot{\theta}_j(t)A_j ,
\end{align}
hence, when considering the compound picture transform 
$\mathscr{P}_N\circ\cdots\circ\mathscr{P}_2\circ\mathscr{P}_1$
on Hamiltonian $\mathcal{H}$, it can obtain the equation
\begin{align}\label{eq13}
	\mathscr{P}_N \circ \cdots \circ \mathscr{P}_2 \circ \mathscr{P}_1(\mathcal{H})=&\mathscr{L}_N \circ \cdots \circ \mathscr{L}_2 \circ \mathscr{L}_1(\mathcal{H})\cr
	&-\Dot{\theta}_1 \mathscr{L}_N \circ \cdots \circ \mathscr{L}_3 \circ \mathscr{L}_2(A_1)   \cr
	&-\Dot{\theta}_2 \mathscr{L}_N \circ \cdots \circ \mathscr{L}_4 \circ \mathscr{L}_3(A_2)  \cr
	&-\cdots \cr
	&-\dot{\theta}_{N-2}(t)\mathscr{L}_N\circ\mathscr{L}_{N-1}(A_{N-2})\cr
	&-\dot{\theta}_{N-1}(t)\mathscr{L}_N(A_{N-1})\cr
	&-\dot{\theta}_N(t)A_N.
\end{align}
It can be observed that Eq.(\ref{eq13}), there are $N$ parameters 
$\{\theta_1, \theta_2,\cdots \theta_n \}$, corresponding to $N$ control parameters $\{ \lambda_1, \lambda_2,\cdots \lambda_n \}$, it opens up the possibility for designing control parameters.

On the other hand, quantum state evolving from the initial state $\psi(t_0)$ is controlled by the evolution operator $U$ and satisfy the Schrödinger equation 
\begin{align}
	 {\rm i}\frac{ \partial}{\partial t} (U\psi(t_0))=\mathcal{H}U\psi(t_0).
\end{align}
Since arbitrariness of $\psi(t_0)$, we can obtain 
\begin{align}
	 {\rm i}\Dot{U}=\mathcal{H}U.
\end{align}
It is equivalent to
\begin{align}\label{eq16}
	0=U^{\dag}\mathcal{H}U-  {\rm i}  U^{\dag}\Dot{U},
\end{align}
comparison of Eq.(\ref{eq13}) and Eq.(\ref{eq16}), when considering 
$\mathscr{P}_N\circ\cdots\circ\mathscr{P}_2\circ\mathscr{P}_1(\mathcal{H})=0$, we can find the relationship between the evolution operator $U$ and the compound picture view transformation. Thus, if the initial time is $t_0$, the evolution operator of the system is
\begin{align}\label{eq17}
	U(t)= {\rm e}^{- {\rm i}\theta_1(t)A_1} {\rm e}^{- {\rm i}\theta_2(t)A_2}\cdots  {\rm e}^{- {\rm i}\theta_N(t)A_N}
	\times  {\rm e}^{ {\rm i}\theta_N(t_0)A_N} {\rm e}^{ {\rm i}\theta_{N-1}(t_0)A_{N-1}}\cdots  {\rm e}^{ {\rm i}\theta_1(t_0)A_1}.
\end{align}
It can derive follow equation from Eq.(\ref{eq16})
\begin{align}\label{eq18}
	\mathscr{L}_N\circ\cdots\circ\mathscr{L}_2\circ\mathscr{L}_1(\mathcal{H})=&\dot{\theta}_1(t)\mathscr{L}_N\circ\cdots\circ\mathscr{L}_3\circ\mathscr{L}_2(A_1) \cr
	&+\dot{\theta}_2(t)\mathscr{L}_N\circ\cdots\circ\mathscr{L}_4\circ\mathscr{L}_3(A_2) \cr
	&+\cdots  \cr
	&+\dot{\theta}_{N-2}(t)\mathscr{L}_N\circ\mathscr{L}_{N-1}(A_{N-2}) \cr
	&+\dot{\theta}_{N-1}(t)\mathscr{L}_N(A_{N-1})\cr
	&+\dot{\theta}_N(t)A_N.
\end{align}
In generally, by designing the parameter $\theta_1, \theta_2 \cdots \theta_N $ within the evolution operator, we can control the evolution of the system's state to achieve the desired state, and the choice of parameter $\theta_1, \theta_2 \cdots \theta_N $ also allows to {inversely
determin} the control parameter $ \lambda_1, \lambda_2,\cdots \lambda_n $. 

\subsection{Interconversion of GHZ state and KLM state}
In this section, we combine effective Hamiltonian Eq. (\ref{eq9})  and Lie-transform-based pulse design to create  control pulses.
We aim to obtain evolution operators in real form and consider them in conjunction with the effective Hamiltonian, thus we choose follow generator
\begin{align}
	A_1=& {\rm i}\ket{\zeta_2}\bra{\zeta_1}- {\rm i}\ket{\zeta_1}\bra{\zeta_2},~~
	A_2= {\rm i}\ket{\zeta_4}\bra{\zeta_3}- {\rm i}\ket{\zeta_3}\bra{\zeta_4},\cr
	A_3=& {\rm i}\ket{\zeta_2}\bra{\zeta_3}- {\rm i}\ket{\zeta_3}\bra{\zeta_2},~~
	A_4= {\rm i}\ket{\zeta_4}\bra{\zeta_1}- {\rm i}\ket{\zeta_1}\bra{\zeta_4},\cr
	A_5=& {\rm i}\ket{\zeta_3}\bra{\zeta_1}- {\rm i}\ket{\zeta_1}\bra{\zeta_3},~~
	A_6= {\rm i}\ket{\zeta_4}\bra{\zeta_2}- {\rm i}\ket{\zeta_2}\bra{\zeta_4},
\end{align}
with $\ket{\zeta_1}=\ket{000}$, $\ket{\zeta_2}=\ket{r00}$, $\ket{\zeta_3}=\ket{rr0}$, $\ket{\zeta_4}=\ket{rrr}$.
Obviously, effective Hamiltonian is
\begin{align}
	H_{ {\rm eff}}=\Omega'_1 A_1 + \Omega'_2 A_3 +\Omega'_3 A_2,
\end{align}
where $\Omega_1= {\rm i}\Omega'_1$,  $\Omega_2=- {\rm i}\Omega'_2$,  $\Omega_3= {\rm i}\Omega'_3$.
From Eq. (\ref{eq18}), we can derive
\begin{align}
	H_{ {\rm eff}}=&~~ \dot{\theta}_1(t)A_1+\dot{\theta}_2(t)\mathscr{L}_1^{-1}A_2+\dot{\theta}_3(t)\mathscr{L}_1^{-1}\mathscr{L}_2^{-1}A_3  \cr
	&+\dot{\theta}_4(t)\mathscr{L}_1^{-1}\mathscr{L}_2^{-1}\mathscr{L}_3^{-1}A_4\cr
	&+\dot{\theta}_5(t)\mathscr{L}_1^{-1}\mathscr{L}_2^{-1}\mathscr{L}_3^{-1}\mathscr{L}_4^{-1}A_5 \cr
	&+\dot{\theta}_6(t)\mathscr{L}_1^{-1}\mathscr{L}_2^{-1}\mathscr{L}_3^{-1}\mathscr{L}_4^{-1}\mathscr{L}_5^{-1}A_6 ,
\end{align}
then, the forms of parameters in control fields can be obtained
\begin{align}
	\Omega'_1=&\Dot{\theta}_1+\Dot{\theta}_5 \cos{\theta_4}\sin{\theta_3}+\Dot{\theta}_6 \cos{\theta_3}\sin{\theta_4},\cr
	\Omega'_2=&\Dot{\theta}_3 \cos{\theta_1} \cos{\theta_2}
	+\Dot{\theta}_4 \sin{\theta_1} \sin{\theta_2}\cr
	&-\Dot{\theta}_5 \cos{\theta_2} \cos{\theta_3}\cos{\theta_4}\sin{\theta_1}\cr
	&-\Dot{\theta}_5 \cos{\theta_1}\sin{\theta_2}\sin{\theta_3}\sin{\theta_4}\cr
	&+\Dot{\theta}_6 \cos{\theta_1} \cos{\theta_3}\cos{\theta_4}\sin{\theta_2}\cr
	&+\Dot{\theta}_6 \cos{\theta_2}\sin{\theta_1}\sin{\theta_3}\sin{\theta_4},\cr
	\Omega'_3=&\Dot{\theta}_2-\Dot{\theta}_5 \cos{\theta_3}\sin{\theta_4}-\Dot{\theta}_6 \cos{\theta_4}\sin{\theta_3},
\end{align}
and constrain conditions
\begin{align}\label{eq23}
	\Dot{\theta}_4=&~ \Dot{\theta}_3 \tan{\theta_1} \tan{\theta_2},\cr
	\Dot{\theta}_5=&~\frac{2\Dot{\theta}_3(\sin\theta_3\sin\theta_4\tan\theta_2-\cos\theta_3\cos\theta_4\tan\theta_1)}{\cos2\theta_3+\cos2\theta_4},\cr  \Dot{\theta}_6=&~\frac{2\Dot{\theta}_3(\cos\theta_3\cos\theta_4\tan\theta_2-\sin\theta_3\sin\theta_4\tan\theta_1)}{\cos2\theta_3+\cos2\theta_4}. \cr
\end{align}
Further, from Eq.(\ref{eq17}), when we set $\theta_k(t_0)=0~(k=1,2\cdots,6)$ in initial time $t_0$, the real evolution operator of effective Hamiltonian (\ref{eq9}) can be written 

	\begin{align}
		&U_{ {\rm eff}}(t) = \cr&\left[
		\begin{array}{cc}
			\cos \theta_1 \cos \theta_4 \cos \theta_5 - \sin \theta_1 \sin \theta_3 \sin \theta_5 & -\cos \theta_3 \cos \theta_6 \sin \theta_1 - \cos \theta_1 \sin \theta_4 \sin \theta_6 \\
			\cos \theta_4 \cos \theta_5  \sin \theta_1 + \cos \theta_1 \sin \theta_3 \sin \theta_5 & \cos \theta_1 \cos \theta_3 \cos \theta_6 - \sin \theta_1 \sin \theta_4 \sin \theta_6 \\
			\cos \theta_2 \cos \theta_3 \sin \theta_5 - \cos \theta_5 \sin \theta_2 \sin \theta_4 & -\cos \theta_2 \cos \theta_6 \sin \theta_3 - \cos \theta_4 \sin \theta_2 \sin \theta_6 \\
			\cos \theta_2 \cos \theta_5 \sin \theta_4 + \cos \theta_3 \sin \theta_2 \sin \theta_5 & \cos \theta_2 \cos \theta_4 \sin \theta_6 - \cos \theta_6 \sin \theta_2 \sin \theta_3 \\
		\end{array}
		\right. \cr
		& \left.
		\begin{array}{cc}
			-\cos \theta_1 \cos \theta_4 \sin \theta_5 - \cos \theta_5 \sin \theta_1 \sin \theta_3 & \cos \theta_3 \sin \theta_1 \sin \theta_6 - \cos \theta_1 \cos \theta_6 \sin \theta_4 \\
			\cos \theta_1 \cos \theta_5 \sin \theta_3 - \cos \theta_4 \sin \theta_1 \sin \theta_5 & -\cos \theta _6 \sin \theta_1 \sin \theta_4 -\cos \theta_1 \cos \theta _3 \sin\theta _6 \\
			\cos \theta_2 \cos \theta_3 \cos \theta_5 + \sin \theta_2  \sin \theta_4\sin \theta_5 & \cos \theta_2  \sin \theta_3 \sin \theta_6 - \cos \theta_4 \cos \theta_6 \sin \theta_2 \\
			-\cos \theta_2 \sin \theta_4 \sin \theta_5 + \cos \theta_3 \cos \theta_5 \sin \theta_2 & \cos \theta_2 \cos \theta_4 \cos \theta_6 +  \sin \theta_2 \sin \theta_3 \sin \theta_6 \\
		\end{array}
		\right] {,}\cr~
	\end{align}

For simplicity, assume that the system initially starts in state $\ket{\zeta_1}$. By controlling the system with the evolution operator $U_{ {\rm eff}}$ of effective Hamiltonian, we can determine the state of the system at any given point in time
\begin{align}\label{eq25}
	U_{ {\rm eff}}\ket{\zeta_1}=\left(
	\begin{array}{c}
		\cos \theta _1 \cos \theta _4 \cos \theta _5-\sin \theta _1 \sin\theta _3 \sin\theta _5 \\
		\cos \theta _1 \sin \theta _3 \sin \theta _5+\sin \theta _1 \cos \theta _4 \cos \theta _5 \\
		\cos\theta _2 \cos \theta_ 3 \sin \theta _5-\sin \theta _2 \sin \theta _4 \cos \theta _5 \\
		\sin \theta _2 \cos \theta _3 \sin \theta _5+\cos \theta _2 \sin \theta _4 \cos \theta _5 \\
	\end{array}
	\right).
\end{align}
Clearly, from Eq.(\ref{eq25}), by determining the boundary conditions for $\theta_k~(k=1,2\cdots,6)$, we can achieve the interconversion between GHZ state and KLM state. We have assumed that the initial state at time $t=t_0$ is $\ket{\zeta_1}=\ket{000}$. Let's further assume that at time $t=t_{{ {\rm ghz}}}$, the system evolves into the GHZ state, and at $t=t_{{ {\rm klm}}}$, it evolves into the KLM state. Specifically speaking, when the boundary conditions satisfy 
\begin{align}\label{eq26}
	\theta_1(t_{{ {\rm ghz}}})&=\theta_2(t_{{ {\rm ghz}}})=\theta_5(t_{{ {\rm ghz}}})=0,
	~\theta_4(t_{{ {\rm ghz}}})=\frac{\pi}{4},
\end{align}
the system will evolve to the $\ket{ {\rm GHZ}}$ state. When the condition is set to 
\begin{align}\label{eq27}
	\theta_1(t_{{ {\rm klm}}})&=-\frac{\pi}{4},~~\theta_2(t_{{ {\rm klm}}})=\frac{\pi}{4},\cr
	\theta_3(t_{{ {\rm klm}}})&=\frac{\pi}{4},~~~~\theta_5(t_{{ {\rm klm}}})=\frac{\pi}{2},
\end{align} 
the system will evolve to the $\ket{ {\rm KLM}}$ state, where $\ket{ {\rm GHZ}}=\left(\ket{\zeta_1}+\ket{\zeta_4}\right)/\sqrt{2}=\left(\ket{000}+\ket{rrr}\right)/\sqrt{2}$, and
$\ket{ {\rm KLM}} = \left( \ket{\zeta_1} + \ket{\zeta_2} + \ket{\zeta_3} + \ket{\zeta_4} \right)/{2} =  \left( \ket{000} + \ket{r00} + \ket{rr0} + \ket{rrr} \right)/{2}$.

\begin{figure}[htbp]
	\centering
	\includegraphics[width=1.0\textwidth]{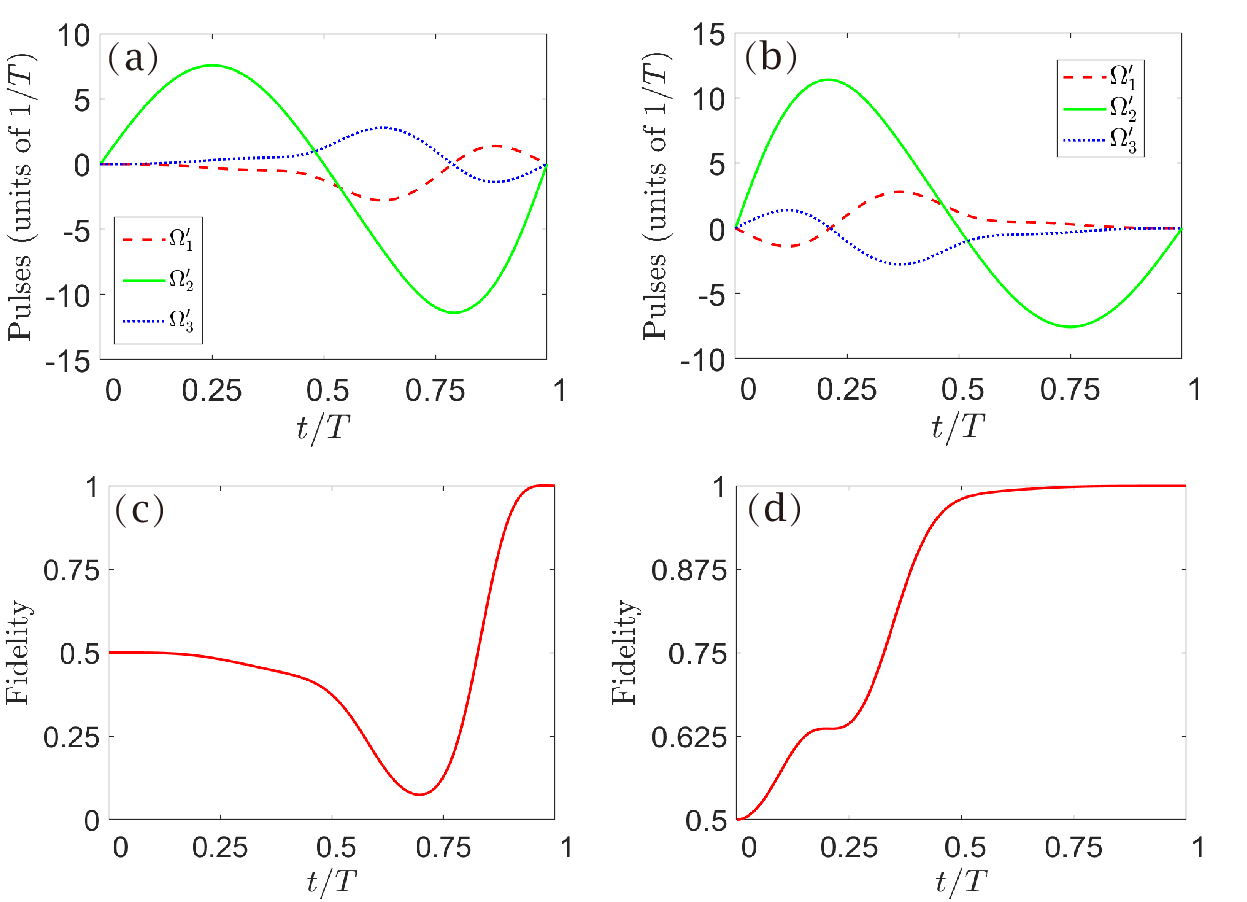}	
	\caption{(a) Rabi frequencies $\Omega_1^{\prime}$, $\Omega_2^{\prime}$, $\Omega_3^{\prime}$(versus $t/T$) for conversion of GHZ to KLM state. (b) Rabi frequencies $\Omega_1^{\prime}$, $\Omega_2^{\prime}$, $\Omega_3^{\prime}$(versus $t/T$) for conversion of KLM to GHZ state. (c) The fidelity of GHZ to KLM state (versus $t/T$) by simulating the Hamiltonian(\ref{eq9}). (d) The fidelity of KLM to GHZ state (versus $t/T$) by simulating the Hamiltonian(\ref{eq9}).}\label{fig2}
\end{figure}

Our focus is on the evolution time between  $t_{{ {\rm ghz}}}$ and $t_{{ {\rm klm}}}$. When we consider GHZ state to KLM state, we can conveniently set $t_{{ {\rm ghz}}}=0$ and $t_{{ {\rm klm}}}=T$.
Through Eq.(\ref{eq26}) and (\ref{eq27}), we can find that $\theta_1$, $\theta_2$, and $\theta_5$ are boundary conditions fixed at both ends, and $\theta_3$ and $\theta_4$ are boundary conditions fixed on one side. The boundary conditions for $\theta_6$ are free at both ends. Eq.(\ref{eq23}) represent the relationship between these parameters. Taking into account the above considerations and no singularities in equation, design parameters can be choose 
\begin{align}
	\theta_1(t)=&-\frac{\pi}{16}\left(1-\cos \frac{\pi t}{T}\right)^2,\cr
	\theta_2(t)=& \frac{\pi}{8}\left( \sin \frac{\pi t}{2T}  \right)^2\left( 1-\cos \frac{\pi t}{T}\right),\cr
	\theta_3(t)=&\frac{\pi}{4} +\mathcal{C}\sin \left(\frac{\pi(t-T)}{T}\right)^2,
\end{align}
where $\mathcal{C}$ is a coefficient to be determined. By numerically calculating Eq.(\ref{eq23}), we obtain $\mathcal{C}=2.4084~$  with boundary conditions $\theta_5(0) = 0$ and $\theta_5(T) = \dfrac{\pi}{2}$. To derive an expression for $\Omega'_1,\Omega'_2,\Omega'_3$, we add $\theta_6(0) = 0$. Through numerical computation, we can then determine $\Omega'_1,\Omega'_2,\Omega'_3$, which exhibits the form shape in Fig.\ref{fig2}(a). Similarly, when we consider KLM state to GHZ state, we only need to let $t_{{ {\rm klm}}}=0$, $t_{{ {\rm ghz}}}=T$. We can easily  {set}
\begin{align}
	\theta_1(t)=&-\frac{\pi}{16}\left(1-\cos \frac{\pi (t-T)}{T}\right)^2,\cr
	\theta_2(t)=& \frac{\pi}{8}\left( \sin \frac{\pi (t-T)}{2T}  \right)^2\left( 1-\cos \frac{\pi (t-T)}{T}\right),\cr
	\theta_3(t)=&\frac{\pi}{4} +\mathcal{C}\sin \left(\frac{\pi t}{T}\right)^2.
\end{align}
And the shape of the control pulse obtained is shown in Fig.\ref{fig2}(b). To demonstrate the effectiveness of our pulse design, we simulate   effective Hamiltonian (\ref{eq9}) and plot the fidelity of the interconversion GHZ state to KLM state as shown in Fig.\ref{fig2}(c) and \ref{fig2}(d), respectively. We can find, this parameter choice has a high fidelity for the effective Hamiltonian. 

\section{Numerical simulation and analyses}\label{sec4}
In this section, we proceed with numerical simulations and analyses. By employing parameters that can be realized experimentally, this study further explores the impact of various decoherence mechanisms on the stability of quantum state evolution. This step aims to verify the effectiveness and robustness of our scheme under realistic experimental conditions.

It is well known that if a unitary transformation is performed from an old picture to a new picture, the quantum state will undergo a phase change. The implementation of the conversion between GHZ state and KLM state takes place under the new picture as defined by Eq. (\ref{eq3}). If discussed in the old picture, a state would experience a phase transformation. Specifically, if a state can be described as $\ket{\psi}$ in the new picture, when transformed back to the original picture, it will becomes to $\mathcal{R} \ket{\psi}$. Although there is a change in phase, once we determine the value of $VT$, we can derive the specific phase of the final state. What we are concerned with is the probability amplitude of the state. Based on this consideration, we define the fidelity at the time $t$ in the new picture as $F=|\bra{\psi} \rho^\prime(t) \ket{\psi}|$, where $\rho(t)$ is the density operator in new picture. So, we define the fidelity of obtaining the KLM state from the GHZ state as $F_1= |\bra{ {\rm KLM}} \rho^\prime(t) \ket{ {\rm KLM}}|$. Similarly, the fidelity of the GHZ state is obtained from the KLM state is $F_2= |\bra{ {\rm GHZ}} \rho^\prime(t)\ket{ {\rm GHZ}}|$. 
\begin{figure}[ht]
	\centering
\includegraphics[width=1.0\textwidth]{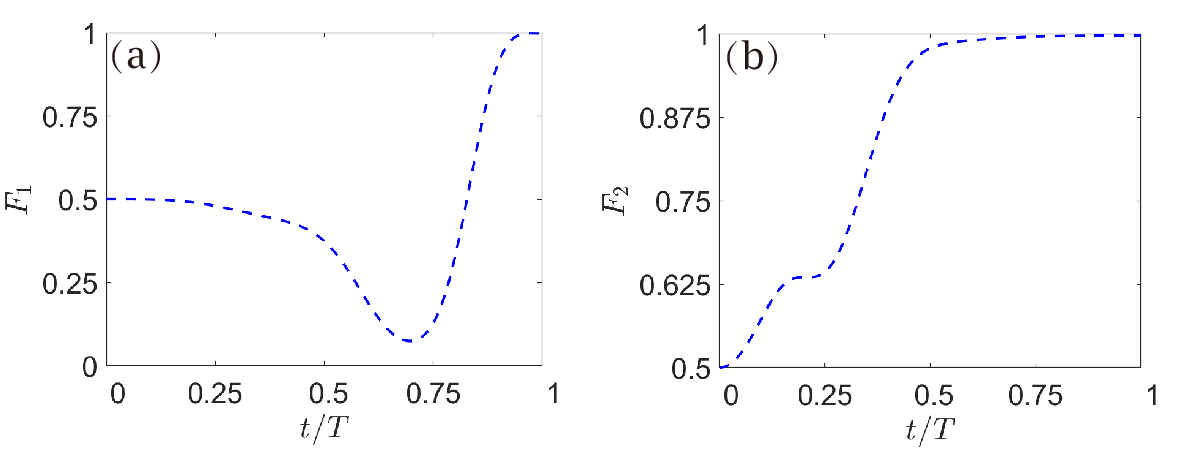}
	\caption{(a) The fidelity $F_1$ versus $t/T$. (b) The fidelity $F_2$ versus $t/T$}\label{F12}
\end{figure}
We performed numerical simulations with the Hamiltonian(\ref{eq1}), the fidelities $F_1$ and $F_2$ are shown in Fig.\ref{F12}, where we have chosen the parameter $V = 100 \pi/T$, which clearly satisfies the $V \gg\mathrm{Max}\{|\Omega_1|, |\Omega_2|, |\Omega_3|\} \approx 11.4 /T$. Numerical simulation results show that the fidelities are more than $99.7\%$ at the moment of $T$. 

Considering the effects of thermal noise, dephasing, and spontaneous emission on the present scheme, we conducted numerical simulations on the fidelities of interconversions between GHZ state and KLM state using the master equation~\cite{PhysRevA.100.012103}
\begin{align}
	\Dot{\rho}=& {-{\rm i}} [H(t),\rho  ]+D_{\rm{deph}}(\rho)+D_{\rm{therm}}(\rho),\cr
	D_{\rm{deph}}=&\sum_{k=1}^3 \left[L_k\rho L^{\dag}_k- \frac{1}{2}\left(L^{\dag}_k L_k\rho+\rho L^{\dag}_k L_k\right)\right],\cr
	D_{\rm{therm}}=&\sum_{k=4}^6 \left \{( \bar{n}+1)\left[L_k\rho L^{\dag}_k- \frac{1}{2}\left(L^{\dag}_k L_k\rho+\rho L^{\dag}_k L_k\right)\right] \right.\cr
	&+\left. \bar{n}\left[L^{\dag}_k\rho L_k- \frac{1}{2}\left(L_k L^{\dag}_k\rho+\rho L_k L^{\dag}_k\right)\right] \right\},
\end{align}
where $\bar{n}$ is the average number of thermal phonons which is set in a scope as 0 $\sim$ 1, when considering the Bose-Einstein distribution, this average phonon number is calculated by $1/( {\rm e}^{\frac{{\hbar \omega}}{k_B \mathcal{T}}}-1)$, which is related to the temperature of the environment $\mathcal{T}$ and the frequency $\omega$ of thermal noise. $L_k (k=1,2,3,4,5,6)$ are the Lindblad operators, which are 
\begin{align}
	L_1=&\sqrt{\gamma_1}\left(\ket{r}_1\bra{r}-\ket{0}_1\bra{0}\right),\cr
	L_2=&\sqrt{\gamma_2}\left(\ket{r}_2\bra{r}-\ket{0}_2\bra{0}\right),\cr
	L_3=&\sqrt{\gamma_3}\left(\ket{r}_3\bra{r}-\ket{0}_3\bra{0}\right), \cr
	L_4=&\sqrt{\Gamma_1}\ket{0}_1\bra{r},\cr
	L_5=&\sqrt{\Gamma_2}\ket{0}_2\bra{r},\cr  
	L_6=&\sqrt{\Gamma_3}\ket{0}_3\bra{r},
\end{align}
$\Gamma_k$ is atomic spontaneous emission rate from $\ket{r}$ to $\ket{0}$, $\gamma_k$ being the dephasing rate of the $k$th atom. For simplicity, here we set $\Gamma_1=\Gamma_2=\Gamma_3=\Gamma$, $\gamma_1=\gamma_2=\gamma_3=\gamma$.
\begin{figure}[htbp]
	\centering

\includegraphics[width=1.0\textwidth]{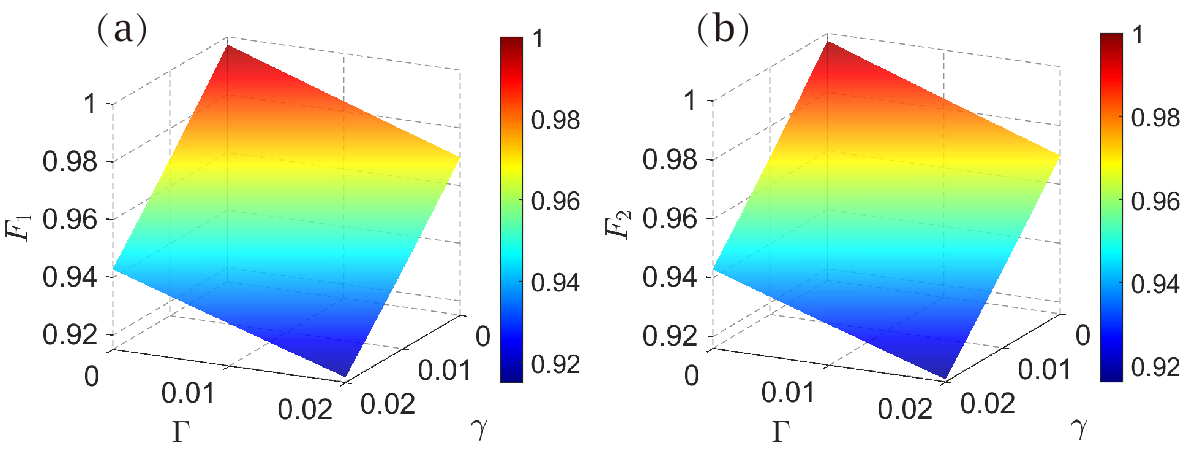}
	\caption{The fidelity(a)$F_1$,(b)$F_2$ versus the decoherence  of spontaneous emission  $\Gamma$ and dephasing  $\gamma$ ( $\Gamma$, $\gamma$ unite of $1/T$).}\label{fig6}
\end{figure}

In Fig.\ref{fig6}, we illustrate the effects of atomic spontaneous emission $\Gamma$ and dephasing rates $\gamma$ on fidelity (note, $\Gamma$, $\gamma$ unite of $1/T$ ), where we have set the average number of thermal phonons $\bar{n}=0$. Through Fig.\ref{fig6}, we can see that this scheme demonstrates robustness against atomic spontaneous emission and dephasing rate.

\begin{figure}[htbp]
	\centering
\includegraphics[width=1.0\textwidth]{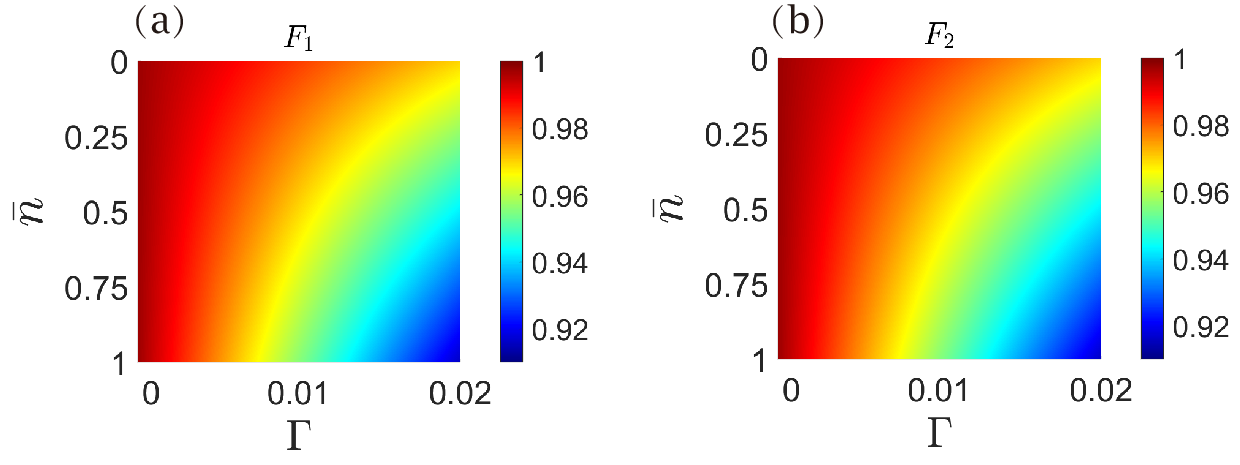}
	\caption{The fidelity(a)$F_1$,(b)$F_2$ versus the  spontaneous emission  $\Gamma$ and and the average number of thermal phonons $\bar{n}$.}\label{fig7}
\end{figure}
Next, we plotted the fidelity as it varies with atomic spontaneous emission and the average number of thermal photons as shown in Fig.\ref{fig7}, even when $\bar{n}=1$ and $\Gamma=0.02$ (unites of $1/T$), the fidelity remains above $91.5\%$. Therefore, we can prove that the scheme is robust against spontaneous emission and the thermal noise.
Here, we consider the total evolutionary time $T=20\mu  {{\rm s}}$, which is still much less than the reported coherence time of Rydberg atom. By considering a group of parameters that are feasible for experiments, $\Gamma=1\rm kHz$, $\gamma=1\rm kHz$ (i.e. 0.02 unites of $1/T$) and experimental temperature is $20 \mu \rm K$~\cite{PhysRevLett.105.193603}, assuming the frequency of thermal noise as $\omega = 2\pi \times 1 \rm MHz$ (i.e. $\bar{n}\sim 0.1$ ). It can obtain $F_1 = 91.16\%$ and $F_2 = 91.17\%$. This indicates the scheme demonstrate robustness against decoherence.

\begin{figure}[htbp]
	\centering
\includegraphics[width=1.0\textwidth]{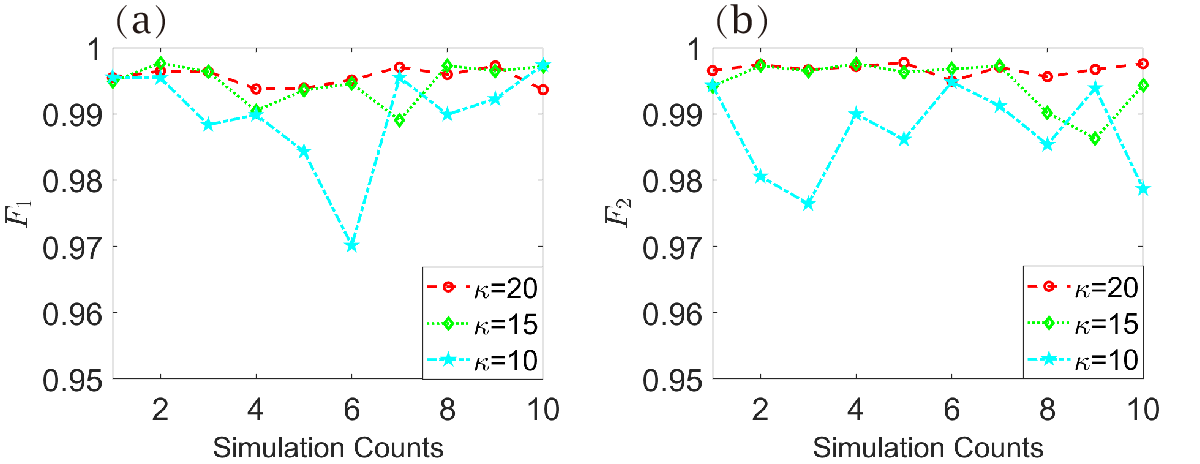}
	\caption{(a) The fidelity $F_1$ of the conversion from GHZ to KLM state  under AWGN versus run number.
		(b) The fidelity $F_2$ of the conversion from KLM to GHZ state  under AWGN versus run number. 
	}\label{fig4}
\end{figure}

In order to further explore the robustness of the protocol under realistic experimental conditions, we will next introduce a factor that is closer to the actual situation, i.e. Gaussian white noise. This noise model will be applied to the control field to simulate the unavoidable environmental disturbances and systematic errors in the experiment. By introducing Gaussian white noise into our control scheme, we evaluate the effect of noise on system evolution and final state fidelities. Then additive white Gaussian noise(AWGN) leads to the noisy control field as 
\begin{align}
	\Omega_k^{no}=\Omega_k +\textrm{Awgn}\left(\Omega_{ {k}}, \kappa\right),
\end{align}
where $\textrm{Awgn}\left(\Omega_{ {k}}, \kappa\right)$ denotes a function-generating AWGN with signal-to-noise ratio (SNR) $\kappa$ for $\Omega_k$. We performed ten separate simulations for each different SNR, the results are shown in the Fig.\ref{fig4}. We can see that even with SNR $\kappa = 10$, the fidelities is still above $97\%$. Therefore, this scheme is robust to Gaussian white noise. 

\section{Summary}\label{sec5}
In summary, we have proposed a scheme to realize the interconversion of GHZ and KLM state of atomic system. We simplify the dynamics by utilizing the energy level shift terms of the Rydberg atomic system, the effective Hamiltonian for the physical model can ultimately be simplified to a four-level Hamiltonian with basis $\{ \ket{000}$, $\ket{r00}$, $\ket{rr0}$, $\ket{rrr}\}$. In this evolutionary space, we implement the interconversion of GHZ and KLM state by Lie-transform-based pulse design. We performed numerical simulations to evaluate the influence of various decoherence factors, such as thermal noise, dephasing and spontaneous emission, the numerical simulation results show that the scheme is robust against various decoherence. In addition, we analyze the effect of Gaussian noise  on the results. In a nutshell, this scheme features deterministic and reversible state conversions, and has high fidelities in both directions. We believe this paper will offer fresh insights and potentially introduce new ideas, contributing to the advancement of quantum information exploration.

\section*{ACKNOWLEDGMENTS}
This work was supported by the Program of the Educational Office of LiaoNing Province of China (Grant Nos. LJKZ1015, LJ2020005, LJKZZ20220120), the Natural Science Foundation of LiaoNing Province (Grant Nos. 2020-BS-234, 2021-MS-317, 2022-MS-372), the Program of Liaoning BaiQianWan Talents Program (Grant No. 2021921096).

\bibliographystyle{apsrev4-2}
\bibliography{refs}

@article{c1,
  title = {Deterministic interconversions between the Greenberger Horne Zeilinger states and the $W$ states by invariant based pulse design},
  author = {Zheng, Ri Hua and Kang, Yi Hao and Ran, Du and Shi, Zhi Cheng and Xia, Yan},
  journal = {Phys. Rev. A},
  volume = {101},
  issue = {1},
  pages = {012345},
  numpages = {11},
  year = {2020},
  month = {Jan},
  publisher = {American Physical Society},
  doi = {10.1103/PhysRevA.101.012345},
  url = {https://link.aps.org/doi/10.1103/PhysRevA.101.012345}
}

@article{Horodecki2009,
  title = {Quantum entanglement},
  author = {Horodecki, Ryszard and Horodecki, Pawe\l{} and Horodecki, Micha\l{} and Horodecki, Karol},
  journal = {Rev. Mod. Phys.},
  volume = {81},
  issue = {2},
  pages = {865  942},
  numpages = {0},
  year = {2009},
  month = {Jun},
  publisher = {American Physical Society},
  doi = {10.1103/RevModPhys.81.865},
  url = {https://link.aps.org/doi/10.1103/RevModPhys.81.865}
}

@article{vedral2014quantum,
  title={Quantum entanglement},
  author={Vedral, Vlatko},
  journal={Nat. Phys.},
  volume={10},
  number={4},
  pages={256  258},
  year={2014},
  publisher={Nature Publishing Group UK London},
  doi = {10.1038/nphys2904},
}

@article{RevModPhys.94.025008,
  title = {Security in quantum cryptography},
  author = {Portmann, Christopher and Renner, Renato},
  journal = {Rev. Mod. Phys.},
  volume = {94},
  issue = {2},
  pages = {025008},
  numpages = {56},
  year = {2022},
  month = {Jun},
  publisher = {American Physical Society},
  doi = {10.1103/RevModPhys.94.025008},
  url = {https://link.aps.org/doi/10.1103/RevModPhys.94.025008}
}

@article{pirandola2020advances,
  title={Advances in quantum cryptography},
  author={Pirandola, Stefano and Andersen, Ulrik L and Banchi, Leonardo and Berta, Mario and Bunandar, Darius and Colbeck, Roger and Englund, Dirk and Gehring, Tobias and Lupo, Cosmo and Ottaviani, Carlo and others},
  journal={Adv. Opt. Photonics},
  volume={12},
  number={4},
  pages={1012  1236},
  year={2020},
  publisher={Optica Publishing Group},
  doi= {10.1364/AOP.361502}
}

@article{bernstein2017post,
  title={Post-quantum cryptography},
  author={Bernstein, Daniel J and Lange, Tanja},
  journal={Nature},
  volume={549},
  number={7671},
  pages={188--194},
  year={2017},
  publisher={Nature Publishing Group UK London},
  doi={10.1038/nature23461}
}

@article{qt1,
  title={Advances in quantum teleportation},
  author={Pirandola, Stefano and Eisert, Jens and Weedbrook, Christian and Furusawa, Akira and Braunstein, Samuel L},
  journal={Nat. Photonics},
  volume={9},
  number={10},
  pages={641  652},
  year={2015},
  publisher={Nature Publishing Group UK London},
  doi= {10.1038/nphoton.2015.154}
}

@article{qt2,
  title={Progress in quantum teleportation},
  author={Hu, Xiao Min and Guo, Yu and Liu, Bi Heng and Li, Chuan Feng and Guo, Guang Can},
  journal={Nature Reviews Physics},
  volume={5},
  number={6},
  pages={339  353},
  year={2023},
  publisher={Nature Publishing Group UK London},
  doi= {10.1038/s42254-023-00588-x}
}

@article{qc1,
  title={Measurement based quantum computation},
  author={Briegel, Hans J and Browne, David E and D{\"u}r, Wolfgang and Raussendorf, Robert and Van den Nest, Maarten},
  journal={Nat. Phys.},
  volume={5},
  number={1},
  pages={19  26},
  year={2009},
  publisher={Nature Publishing Group UK London},
  doi= {10.1038/nphys1157}
}

@article{qc3,
  title = {Adiabatic quantum computation},
  author = {Albash, Tameem and Lidar, Daniel A.},
  journal = {Rev. Mod. Phys.},
  volume = {90},
  issue = {1},
  pages = {015002},
  numpages = {64},
  year = {2018},
  month = {Jan},
  publisher = {American Physical Society},
  doi = {10.1103/RevModPhys.90.015002},
  url = {https://link.aps.org/doi/10.1103/RevModPhys.90.015002}
}

@article{qd1,
  title = {The security of practical quantum key distribution},
  author = {Scarani, Valerio and Bechmann Pasquinucci, Helle and Cerf, Nicolas J. and Du\ifmmode \check{s}\else \v{s}\fi{}ek, Miloslav and L\"utkenhaus, Norbert and Peev, Momtchil},
  journal = {Rev. Mod. Phys.},
  volume = {81},
  issue = {3},
  pages = {1301  1350},
  numpages = {0},
  year = {2009},
  month = {Sep},
  publisher = {American Physical Society},
  doi = {10.1103/RevModPhys.81.1301},
  url = {https://link.aps.org/doi/10.1103/RevModPhys.81.1301}
}

@article{qd2,
  title = {Secure quantum key distribution with realistic devices},
  author = {Xu, Feihu and Ma, Xiongfeng and Zhang, Qiang and Lo, Hoi Kwong and Pan, Jian Wei},
  journal = {Rev. Mod. Phys.},
  volume = {92},
  issue = {2},
  pages = {025002},
  numpages = {60},
  year = {2020},
  month = {May},
  publisher = {American Physical Society},
  doi = {10.1103/RevModPhys.92.025002},
  url = {https://link.aps.org/doi/10.1103/RevModPhys.92.025002}
}

@article{ji2017conversion,
  title={Conversion of entangled states with nitrogen vacancy centers coupled to microtoroidal resonators},
  author={Ji, Y Q and Shao, X Q and Yi, X X},
  journal={Opt. Express},
  volume={25},
  number={14},
  pages={15806  15817},
  year={2017},
  publisher={Optica Publishing Group},
doi= {10.1364/OE.25.015806}  
}

@article{shen2019conversion,
  title={Conversion of knill  laflamme  milburn entanglement to greenberger  horne  zeilinger entanglement},
  author={Shen, Cai Peng and Xiu, Xiao Ming and Dong, Li and Zhu, Xiao Yu and Chen, Li and Liang, Erjun and Yan, Lei Lei and Su, Shi Lei},
  journal={Ann. Phys.},
  volume={531},
  number={12},
  pages={1900160},
  year={2019},
  publisher={Wiley Online Library},
  doi= {10.1002/andp.201900160}
}

@article{Song2012,
  title = {Direct conversion of a four atom $W$ state to a Greenberger Horne Zeilinger state via a dissipative process},
  author = {Song, Jie and Sun, Xiu Dong and Mu, Qing Xia and Zhang, Ling Li and Xia, Yan and Song, He Shan},
  journal = {Phys. Rev. A},
  volume = {88},
  issue = {2},
  pages = {024305},
  numpages = {4},
  year = {2013},
  month = {Aug},
  publisher = {American Physical Society},
  doi = {10.1103/PhysRevA.88.024305},
  url = {https://link.aps.org/doi/10.1103/PhysRevA.88.024305}
}

@article{shao2023,
  title = {High Fidelity Interconversion between Greenberger Horne Zeilinger and $W$ States through Floquet Lindblad Engineering in Rydberg Atom Arrays},
  author = {Shao, X. Q. and Liu, F. and Xue, X. W. and Mu, W. L. and Li, Weibin},
  journal = {Phys. Rev. Appl.},
  volume = {20},
  issue = {1},
  pages = {014014},
  numpages = {19},
  year = {2023},
  month = {Jul},
  publisher = {American Physical Society},
  doi = {10.1103/PhysRevApplied.20.014014},
  url = {https://link.aps.org/doi/10.1103/PhysRevApplied.20.014014}
}

@article{zhang2024fast,
  title={Fast Preparation of W States by Designing the Evolution Operators with Rydberg Superatom},
  author={Zhang, Wei Yue and Wang, Cai Qing and Ji, Yan Qiang and Shao, Qiu Ping and Wang, Ji Ping and Wang, Jie and Yang, Li Ping and Dong, Li and Xiu, Xiao Ming},
  journal={Advanced Quantum Technologies},
  volume={7},
  number={1},
  pages={2300140},
  year={2024},
  publisher={Wiley Online Library},
  doi= {10.1002/qute.202300140}
}

@article{Kang2019,
  title = {Deterministic conversions between Greenberger Horne Zeilinger states and $W$ states of spin qubits via Lie transform based inverse Hamiltonian engineering},
  author = {Kang, Yi Hao and Shi, Zhi Cheng and Huang, Bi Hua and Song, Jie and Xia, Yan},
  journal = {Phys. Rev. A},
  volume = {100},
  issue = {1},
  pages = {012332},
  numpages = {12},
  year = {2019},
  month = {Jul},
  publisher = {American Physical Society},
  doi = {10.1103/PhysRevA.100.012332},
  url = {https://link.aps.org/doi/10.1103/PhysRevA.100.012332}
}

@article{Ji2019,
  title = {Fast conversion of Dicke states $|{D}_{n}^{(2)}\ensuremath{\rangle}$ to $|{D}_{n+1}^{(2)}\ensuremath{\rangle}$ by transitionless quantum driving},
  author = {Ji, Yan Qiang and Liu, Y. L. and Zhou, S. J. and Xiu, X. M. and Dong, L. and Dong, H. K. and Gao, Y. J. and Yi, X. X.},
  journal = {Phys. Rev. A},
  volume = {99},
  issue = {2},
  pages = {023808},
  numpages = {9},
  year = {2019},
  month = {Feb},
  publisher = {American Physical Society},
  doi = {10.1103/PhysRevA.99.023808},
  url = {https://link.aps.org/doi/10.1103/PhysRevA.99.023808}
}

@article{Gujarati2018,
  title = {Rydberg-atom based creation of an $N$ particle Greenberger Horne Zeilinger state using stimulated Raman adiabatic passage},
  author = {Gujarati, Tanvi P.},
  journal = {Phys. Rev. A},
  volume = {98},
  issue = {6},
  pages = {062326},
  numpages = {7},
  year = {2018},
  month = {Dec},
  publisher = {American Physical Society},
  doi = {10.1103/PhysRevA.98.062326},
  url = {https://link.aps.org/doi/10.1103/PhysRevA.98.062326}
}

@article{wu2017superadiabatic,
  title={Superadiabatic scheme for optimizing the fast generation of tree type 3D entanglement},
  author={Wu, Jin Lei and Su, Shi Lei and Ji, Xin and Zhang, Shou},
  journal={Ann. Phys.},
  volume={386},
  pages={34  43},
  year={2017},
  publisher={Elsevier},
  doi= {10.1016/j.aop.2017.08.014}
}

@article{urban2009observation,
  title={Observation of Rydberg blockade between two atoms},
  author={Urban, E and Johnson, Todd A and Henage, T and Isenhower, L and Yavuz, DD and Walker, TG and Saffman, M},
  journal={Nat. Phys.},
  volume={5},
  number={2},
  pages={110  114},
  year={2009},
  publisher={Nature Publishing Group UK London},
  doi= {10.1038/nphys1178}
}

@article{gaetan2009observation,
  title={Observation of collective excitation of two individual atoms in the Rydberg blockade regime},
  author={Ga{\"e}tan, Alpha and Miroshnychenko, Yevhen and Wilk, Tatjana and Chotia, Amodsen and Viteau, Matthieu and Comparat, Daniel and Pillet, Pierre and Browaeys, Antoine and Grangier, Philippe},
  journal={Nat. Phys.},
  volume={5},
  number={2},
  pages={115  118},
  year={2009},
  publisher={Nature Publishing Group UK London},
  doi= {10.1038/nphys1183}
  
}

@article{Saffman2009,
  title = {Quantum information with Rydberg atoms},
  author = {Saffman, M. and Walker, T. G. and M\o{}lmer, K.},
  journal = {Rev. Mod. Phys.},
  volume = {82},
  issue = {3},
  pages = {2313  2363},
  numpages = {0},
  year = {2010},
  month = {Aug},
  publisher = {American Physical Society},
  doi = {10.1103/RevModPhys.82.2313},
  url = {https://link.aps.org/doi/10.1103/RevModPhys.82.2313}
}

@article{PhysRevLett.104.013001,
  title = {Evidence of Antiblockade in an Ultracold Rydberg Gas},
  author = {Amthor, Thomas and Giese, Christian and Hofmann, Christoph S. and Weidem\"uller, Matthias},
  journal = {Phys. Rev. Lett.},
  volume = {104},
  issue = {1},
  pages = {013001},
  numpages = {4},
  year = {2010},
  month = {Jan},
  publisher = {American Physical Society},
  doi = {10.1103/PhysRevLett.104.013001},
  url = {https://link.aps.org/doi/10.1103/PhysRevLett.104.013001}
}

@article{su2020rydberg,
  title={Rydberg antiblockade regimes: dynamics and applications},
  author={Su, S L and Guo, F Q and Wu, J L and Jin, Z and Shao, X Q and Zhang, S},
  journal={Europhys. Lett.},
  volume={131},
  number={5},
  pages={53001},
  year={2020},
  publisher={IOP Publishing},
  doi={10.1209/0295-5075/131/53001}
}

@article{Su2021Dipole,
  title = {Dipole dipole interaction  driven antiblockade of two Rydberg atoms},
  author = {Su, Shi Lei and Li, Weibin},
  journal = {Phys. Rev. A},
  volume = {104},
  issue = {3},
  pages = {033716},
  numpages = {13},
  year = {2021},
  month = {Sep},
  publisher = {American Physical Society},
  doi = {10.1103/PhysRevA.104.033716},
  url = {https://link.aps.org/doi/10.1103/PhysRevA.104.033716}
}

@article{Shao2017,
  title = {Ground state blockade of Rydberg atoms and application in entanglement generation},
  author = {Shao,  Xiao Qiang and Li, D. X. and Ji, Yan Qiang and Wu, J. H. and Yi, X. X.},
  journal = {Phys. Rev. A},
  volume = {96},
  issue = {1},
  pages = {012328},
  numpages = {9},
  year = {2017},
  month = {Jul},
  publisher = {American Physical Society},
  doi = {10.1103/PhysRevA.96.012328},
  url = {https://link.aps.org/doi/10.1103/PhysRevA.96.012328}
}

@article{liu2022,
  title = {Generation of cat states by a weak parametric drive and a transitionless tracking algorithm},
  author = {Liu, Shuai and Chen, Ye Hong and Wang, Yu and Kang, Yi Hao and Shi, Zhi Cheng and Song, Jie and Xia, Yan},
  journal = {Phys. Rev. A},
  volume = {106},
  issue = {4},
  pages = {042430},
  numpages = {11},
  year = {2022},
  month = {Oct},
  publisher = {American Physical Society},
  doi = {10.1103/PhysRevA.106.042430},
  url = {https://link.aps.org/doi/10.1103/PhysRevA.106.042430}
}

@article{kang2019pulse,
  title = {Pulse design for multilevel systems by utilizing Lie transforms},
  author = {Kang, Yi Hao and Chen, Ye Hong and Shi, Zhi Cheng and Huang, Bi Hua and Song, Jie and Xia, Yan},
  journal = {Phys. Rev. A},
  volume = {97},
  issue = {3},
  pages = {033407},
  numpages = {13},
  year = {2018},
  month = {Mar},
  publisher = {American Physical Society},
  doi = {10.1103/PhysRevA.97.033407},
  url = {https://link.aps.org/doi/10.1103/PhysRevA.97.033407}
}

@article{KLM,
  title={A scheme for efficient quantum computation with linear optics},
  author={Knill, Emanuel and Laflamme, Raymond and Milburn, Gerald J},
  journal={Nature},
  volume={409},
  number={6816},
  pages={46  52},
  year={2001},
  publisher={Nature Publishing Group UK London},
  doi= {10.1038/35051009}
}

@article{cheng2012generation,
  title={Generation of two atom Knill  Laflamme  Milburn states with cavity quantum electrodynamics},
  author={Cheng, Liu Yong and Wang, Hong Fu and Zhang, Shou and Yeon, Kyu Hwang},
  journal={JOSA B},
  volume={29},
  number={7},
  pages={1584  1588},
  year={2012},
  publisher={Optica Publishing Group},
  doi= {10.1364/JOSAB.29.001584}
}

@article{ji2020preparation,
  title={Preparation of entangled states in multiple cavities},
  author={Ji, Y. Q. and Li, H and Liu, Y. L. and Zhang, D. W.  and Zhou, X. J. and Xiao, R. J. and Dong, L. and Xiu, X. M.},
  journal={Laser Phys. Lett.},
  volume={17},
  number={8},
  pages={085202},
  year={2020},
  publisher={IOP Publishing},
  doi={10.1088/1612-202X/ab96c9}
}

@article{li2018engineering,
  title={Engineering steady Knill Laflamme Milburn state of Rydberg atoms by dissipation},
  author={Li, Dong Xiao and Shao, Xiao Qiang and Wu, Jin Hui and Yi, XX and Zheng, Tai Yu},
  journal={Opt. Express},
  volume={26},
  number={3},
  pages={2292  2302},
  year={2018},
  publisher={Optica Publishing Group},
  doi= {10.1364/OE.26.002292}
}

@article{shen2018multiphoton,
  title={Multiphoton Knill Laflamme Milburn states generated by nonlinear optics},
  author={Shen, Cai Peng and Gu, Xiao Fei and Guo, Qi and Zhu, Xiao Yu and Su, Shi Lei and Liang, Erjun},
  journal={JOSA B},
  volume={35},
  number={4},
  pages={694  701},
  year={2018},
  publisher={Optica Publishing Group},
  doi= {10.1364/JOSAB.35.000694}
}

@article{zheng2021Fast,
  title = {Fast and dephasing tolerant preparation of steady Knill Laflamme Milburn states via dissipative Rydberg pumping},
  author = {Zheng, Ri Hua and Xiao, Yang and Su, S. L. and Chen, Ye Hong and Shi, Zhi Cheng and Song, Jie and Xia, Yan and Zheng, Shi Biao},
  journal = {Phys. Rev. A},
  volume = {103},
  issue = {5},
  pages = {052402},
  numpages = {10},
  year = {2021},
  month = {May},
  publisher = {American Physical Society},
  doi = {10.1103/PhysRevA.103.052402},
  url = {https://link.aps.org/doi/10.1103/PhysRevA.103.052402}
}

@article{Lemr2010Experimental,
  title = {Experimental preparation of two photon Knill Laflamme Milburn states},
  author = {Lemr, Karel and \ifmmode \check{C}\else \v{C}\fi{}ernoch, Anton\'{\i}n and Soubusta, Jan and Fiur\'a\ifmmode \check{s}\else \v{s}\fi{}ek, Jarom\'{\i}r},
  journal = {Phys. Rev. A},
  volume = {81},
  issue = {1},
  pages = {012321},
  numpages = {4},
  year = {2010},
  month = {Jan},
  publisher = {American Physical Society},
  doi = {10.1103/PhysRevA.81.012321},
  url = {https://link.aps.org/doi/10.1103/PhysRevA.81.012321}
}

@article{okamoto2011realization,
  title={Realization of a Knill Laflamme Milburn controlled NOT photonic quantum circuit combining effective optical nonlinearities},
  author={Okamoto, Ryo and O’Brien, Jeremy L and Hofmann, Holger F and Takeuchi, Shigeki},
  journal={Proc. Natl. Acad. Sci.},
  volume={108},
  number={25},
  pages={10067  10071},
  year={2011},
  publisher={National Acad Sciences},
  doi= {10.1073/pnas.1018839108}
}

@article{PhysRevLett.105.193603,
  title = {Cooperative Atom Light Interaction in a Blockaded Rydberg Ensemble},
  author = {Pritchard, J. D. and Maxwell, D. and Gauguet, A. and Weatherill, K. J. and Jones, M. P. A. and Adams, C. S.},
  journal = {Phys. Rev. Lett.},
  volume = {105},
  issue = {19},
  pages = {193603},
  numpages = {4},
  year = {2010},
  month = {Nov},
  publisher = {American Physical Society},
  doi = {10.1103/PhysRevLett.105.193603},
  url = {https://link.aps.org/doi/10.1103/PhysRevLett.105.193603}
}

@article{PhysRevA.100.012103,
  title = {Pulse engineering for population control under dephasing and dissipation},
  author = {Medina, I. and Semi\~ao, F. L.},
  journal = {Phys. Rev. A},
  volume = {100},
  issue = {1},
  pages = {012103},
  numpages = {7},
  year = {2019},
  month = {Jul},
  publisher = {American Physical Society},
  doi = {10.1103/PhysRevA.100.012103},
  url = {https://link.aps.org/doi/10.1103/PhysRevA.100.012103}
}

@article{shiQuantumLogicEntanglement2022,
 title = {Quantum Logic and Entanglement by Neutral {{Rydberg}} Atoms},
  author = {Shi, Xiao-Feng},
  date = {2022-04-01},
  year={2022},
  journal = {Quantum Sci. Technol.},
  volume = {7},
  number = {2},
  pages = {023002},
  issn = {2058-9565},
  doi = {10.1088/2058-9565/ac18b8}
}

@article{liu2023fast,
  title={Fast Conversion of Three-Particle Dicke States to Four-Particle Dicke States with Rydberg Superatoms},
  author={Liu, Ying Li and Ji, Yan Qiang and Han, Xue and Cui, Wen Xue and Zhang, Shou and Wang, Hong Fu},
  journal={Adv. Quantum Technol.},
  volume={6},
  number={4},
  pages={2200173},
  year={2023},
  publisher={Wiley Online Library},
  doi={10.1002/qute.202200173}
}

@article{PhysRevA.103.032427,
  title = {Conversion from $W$ to Greenberger-Horne-Zeilinger states in the Rydberg-blockade regime of neutral-atom systems: Dynamical-symmetry-based approach},
  author = {Haase, Thorsten and Alber, Gernot and Stojanovi\ifmmode \acute{c}\else \'{c}\fi{}, Vladimir M.},
  journal = {Phys. Rev. A},
  volume = {103},
  issue = {3},
  pages = {032427},
  numpages = {10},
  year = {2021},
  month = {Mar},
  publisher = {American Physical Society},
  doi = {10.1103/PhysRevA.103.032427},
  url = {https://link.aps.org/doi/10.1103/PhysRevA.103.032427}
}

@article{ghz1,
  title={Shortcuts to adiabatic passage for fast generation of Greenberger-Horne-Zeilinger states by transitionless quantum driving},
  author={Chen, Ye Hong and Xia, Yan and Song, Jie and Chen, Qing Qin},
  journal={Sci. Rep.},
  volume={5},
  number={1},
  pages={15616},
  year={2015},
  publisher={Nature Publishing Group UK London},
  doi={10.1038/srep15616}
}

@article{ghz2,
  title={Adiabatic passage for one-step generation of n-qubit Greenberger--Horne--Zeilinger states of superconducting qubits via quantum Zeno dynamics},
  author={Wu, Jin Lei and Song, Chong and Xu, Jing and Yu, Lin and Ji, Xin and Zhang, Shou},
  journal={Quantum Inf. Process.},
  volume={15},
  pages={3663--3675},
  year={2016},
  publisher={Springer},
  doi={10.1007/s11128-016-1366-0}
}

@article{klm1,
  title={Engineering Knill--Laflamme--Milburn Entanglement via Dissipation and Coherent Population Trapping in Rydberg Atoms},
  author={Li, Rui and He, Shuang and Meng, Zhi-Jun and Jin, Zhao and Gong, Wei-Jiang},
  journal={Chin. Phys. Lett.},
  volume={40},
  number={6},
  pages={060302},
  year={2023},
  publisher={IOP Publishing},
  doi={10.1088/0256-307X/40/6/060302}
}

@article{monroe2002quantum,
  title={Quantum information processing with atoms and photons},
  author={Monroe, Chris},
  journal={Nature},
  volume={416},
  number={6877},
  pages={238--246},
  year={2002},
  publisher={Nature Publishing Group UK London}
}

@article{blais2020quantum,
  title={Quantum information processing and quantum optics with circuit quantum electrodynamics},
  author={Blais, Alexandre and Girvin, Steven M and Oliver, William D},
  journal={Nat. Phys.},
  volume={16},
  number={3},
  pages={247--256},
  year={2020},
  publisher={Nature Publishing Group UK London}
}

@article{xavier2020quantum,
  title={Quantum information processing with space-division multiplexing optical fibres},
  author={Xavier, Guilherme B and Lima, Gustavo},
  journal={Communications Physics},
  volume={3},
  number={1},
  pages={9},
  year={2020},
  publisher={Nature Publishing Group UK London}
}

@article{popescu2007knill,
  title={Knill-Laflamme-Milburn quantum computation with bosonic atoms},
  author={Popescu, Sandu},
  journal={Phys. Rev. Lett.},
  volume={99},
  number={13},
  pages={130503},
  year={2007},
  publisher={APS}
}

@article{shao2020s,
  title = {Selective Rydberg pumping via strong dipole blockade},
  author = {Shao, Xiao-Qiang},
  journal = {Phys. Rev. A},
  volume = {102},
  issue = {5},
  pages = {053118},
  numpages = {11},
  year = {2020},
  month = {Nov},
  publisher = {American Physical Society},
  doi = {10.1103/PhysRevA.102.053118},
  url = {https://link.aps.org/doi/10.1103/PhysRevA.102.053118}
}

@article{nauth2022quantum,
  title={Quantum-brachistochrone approach to the conversion from W to Greenberger-Horne-Zeilinger states for Rydberg-atom qubits},
  author={Nauth, Julian K and Stojanovi{\'c}, Vladimir M},
  journal={Phys. Rev. A},
  volume={106},
  number={3},
  pages={032605},
  year={2022},
  publisher={APS}
}

@article{santos2016shortcut,
  title={Shortcut to adiabatic gate teleportation},
  author={Santos, Alan C and Silva, Raphael D and Sarandy, Marcelo S},
  journal={Phys. Rev. A},
  volume={93},
  number={1},
  pages={012311},
  year={2016},
  publisher={APS}
}

@article{martinez2014shortcuts,
  title={Shortcuts to adiabaticity in three-level systems using Lie transforms},
  author={Mart{\'\i}nez-Garaot, Sof{\'\i}a and Torrontegui, E and Chen, Xi and Muga, Juan Gonzalo},
  journal={Phys. Rev. A},
  volume={89},
  number={5},
  pages={053408},
  year={2014},
  publisher={APS}
}

@article{shor2000simple,
  title={Simple proof of security of the BB84 quantum key distribution protocol},
  author={Shor, Peter W and Preskill, John},
  journal={Phys. Rev. Lett.},
  volume={85},
  number={2},
  pages={441},
  year={2000},
  publisher={APS}
}

@article{yang2024fast,
  title={Fast generation of GHZ state with Rydberg superatom by transitionless quantum driving},
  author={Yang, LP and Wang, JP and Ji, YQ and Wang, J and Zhang, ZM and Liu, YL and Dong, L and Xiu, XM},
  journal={The European Physical Journal Plus},
  volume={139},
  number={7},
  pages={1--12},
  year={2024},
  publisher={Springer}
}

@article{mintert2005measures,
  title={Measures and dynamics of entangled states},
  author={Mintert, Florian and Carvalho, Andr{\'e} RR and Ku{\'s}, Marek and Buchleitner, Andreas},
  journal={Phys. Rep.},
  volume={415},
  number={4},
  pages={207--259},
  year={2005},
  publisher={Elsevier}
}

@article{2017singlet,
  title={Singlet-State Preparation for Three $\Lambda$-type Atoms with Rydberg Blockade Mechanism},
  author={Rong-Can Yang, Xiu Lin, Hong-Yu Liu},
  journal={Commun. Theor. Phys.},
  volume={67},
  number={8},
  pages={215--219},
  year={2017}
}

@article{adams2019rydberg,
  title={Rydberg atom quantum technologies},
  author={Adams, Charles S and Pritchard, Jonathan D and Shaffer, James P},
  journal={J. Phys. B: At., Mol. Opt. Phys.},
  volume={53},
  number={1},
  pages={012002},
  year={2019},
  publisher={IOP Publishing}
}

@article{chen2018fast,
  title={Fast creation of a three-atom singlet state with a dissipative mechanism and Rydberg blockade},
  author={Chen, Xiang and Lin, Gong-Wei and Xie, Hong and Shang, Xiao and Ye, Ming-Yong and Lin, Xiu-Min},
  journal={Phys. Rev. A},
  volume={98},
  number={4},
  pages={042335},
  year={2018},
  publisher={APS}
}

@article{yu2022multiqubit,
  title={Multiqubit Toffoli gates and optimal geometry with Rydberg atoms},
  author={Yu, Dongmin and Wang, Han and Liu, Jin-Ming and Su, Shi-Lei and Qian, Jing and Zhang, Weiping},
  journal={Phys. Rev. Appl},
  volume={18},
  number={3},
  pages={034072},
  year={2022},
  publisher={APS}
}


\end{document}